\documentstyle[12pt,aaspp4]{article}

\eqsecnum

\newcommand{\be}{\begin{equation}}
\newcommand{\ee}{\end{equation}}
\newcommand{\beqa}{\begin{eqnarray}}
\newcommand{\eeqa}{\end{eqnarray}}
\newcommand{\beqan}{\begin{eqnarray*}}
\newcommand{\eeqan}{\end{eqnarray*}}

\begin{document}

\title{Rethinking the Model of Moving Cosmic Strings \\
Producing Anisotropy in the Microwave Background}
\author{Francine R. Marleau, Charles C. Dyer and Jody H. Palmer}
\affil{Department of Astronomy, University of Toronto, \\
60 St. George Street, Toronto, Ontario, Canada M5S 1A7}

\begin{abstract}
We re-analyze the issue of redshifts induced by a moving cosmic string by
looking at moving sources and observers on a conical spacetime in a fully
relativistic approach.  By replacing the concept of a moving spacetime
with the more clearly defined concept of moving sources and observers in the
string spacetime, we show that there is no
effect: the only redshift is a Doppler shift due to the motion of the
source or observer.
\end{abstract}


\section{Introduction}

Cosmic strings are topological defects that might be created
in the early universe during the phase transition/symmetry breaking
that occurs in the classical field theory describing the universal
forces at that time (Kibble 1981; Vilenkin 1985).   These
strings evolve into infinitely long vortices or small loops
that decay over time by radiating gravitational waves.  According
to the literature (Vilenkin 1985),
infinitely long cosmic strings have a gravitational
field associated with a conical Minkowski spacetime and have
acquired relativistic motion after their formation.
It is commonly postulated that cosmic strings might provide the seeds for
the formation of inhomogeneities in the universe, and that their presence
will be revealed by inhomogeneities in the cosmic microwave background
(CMB)
(Kaiser \& Stebbins 1984; Stebbins 1988).  Because of the similarity of a
conical spacetime to a topologically flat Minkowski spacetime, the effects
of strings on the microwave background have been treated in a
nonrelativistic fashion in the literature.  Here, we use a relativistic
formulation to analyze physics on a string spacetime
and to comment on the effects of a string on the microwave background.

When modeling the presence of physical strings in the universe, one must
cope with a particularly difficult problem: a string is a topological
formation and there is no accepted mechanism for the topological evolution
of a spacetime.  Therefore, concepts such as the creation of a string
spacetime,
and the motion of strings through the universe must be dealt with through
simplified models.  The greatest simplification comes by studying the
spacetime associated  with a single infinitely long string.  This is not
without its drawbacks.  In particular, it has been shown that an infinitely
long string is not embeddable in a homogeneous and isotropic cosmology
(Dyer, Oattes \& Starkman 1988), and thus is of limited use when discussing
cosmological issues.

The spacetime associated with a self-gravitating cosmic string was
introduced in 1981 by Vilenkin.  In the weak-field limit, he found that
the spacetime was identically conical Minkowski.
Using numerical methods, it has been recently shown (Dyer \& Marleau
1995; Laguna-Castillo \& Matzner 1987) that static cosmic strings described
as infinite cylinders in a U(1) model are self-gravitating objects with a
spacetime that is not conical Minkowski.
These two results are reconciled by realizing that the spacetime is curved
near the string, and asymptotically approaches conical Minkowski.  By
recalculating Vilenkin's result, we find that the cause of the discrepancy
is an invalid coordinate transformation in going from his equation (27) to
his equation (31).  This can
be seen by calculating the curvature associated with the two
metrics.  Before the transformation, the Riemann curvature tensor is
non vanishing; after the  transformation, the Riemann tensor is zero.
Fortunately, the impact of this error has been slight because the field
is asymptotically conical Minkowski.
While more complete solutions have been
investigated, the conical solution is commonly used in the analysis of
strings in cosmology.

The simplicity of the conical solution has lead to a
non-relativistic treatment of the lensing and redshift properties
of cosmic strings.  In particular, the effects of a moving string have been
modeled by moving a wedge representing the deficit angle with respect
to otherwise ``stationary'' observers.  This model is difficult to apply
self consistently since it is not known how to move a
string and calculate the back-reaction of the rest of the universe; moving
the string is not equivalent to moving the source and observer.
Kaiser and Stebbins use this picture of a conical spacetime as
``Minkowski minus a wedge" to calculate the effect of a moving string on
temperature fluctuations of the microwave background.
In this approach the spacetime is modeled by imagining the spacelike
slices with
constant $z$ as flat pieces of paper with a wedge (with apex at the string)
removed.  Then a source of radiation and an observer are set on the paper
such that they are not moving with respect to each other.  Finally, the
motion of the string with respect to the source and observer is introduced
by dragging the wedge across the paper between the source and observer.

Kaiser and Stebbins found that the results of this procedure were:
First, the observer sees one image, then as the string moves across the sky
between the source and observer, the observer sees two images.
Second, the observer views the source with no redshift before
the string passes, and with a negative redshift after the string has
passed.
In addition to these results, Kaiser and Stebbins applied the ``moving lens
method" from the theory of gravitational lenses (Birkinshaw \& Gull 1983)
to find the perturbation of the microwave background caused by a
moving string.  They found that there should be a jump discontinuity in the
observed temperature of the microwave background at the location of the
string.

In this paper we re-analyze these results through modeling a moving string
by placing a moving source and a moving observer on a conical spacetime.
Now the motion of the string is modeled by the motion of the source and
observer which is understood in terms of the equations of timelike
geodesics on the conical spacetime.  The multiple imaging characteristics
of strings are analyzed through the equations for null geodesics, and the
redshift of the microwave background is calculated directly from the
variation of $u_ak^a$ as in the usual definition of frequency and
redshift in general relativity.

\section{Geodesics on a Conical Spacetime}

The motions of the source and (nonaccelerated) observer on a conical
spacetime are governed by the geodesic equations.  For a conical spacetime
the metric has the form
\be
ds^2 = -dt^2 + dr^2 + \alpha^2 r^2 d\phi^2 + dz^2
\ee
in the usual coordinates $(t,r,\phi,z)$.
The conical nature of this spacetime is due to the deficit angle
$2\pi(1-\alpha)$.  A value of $\alpha = 1$ gives the usual Minkowski
spacetime.

The null geodesics on a conical spacetime are very similar to the timelike
geodesics.
The geodesic tangent vector has the components:
\be
\left( -A, \pm\sqrt{C - \frac{B^2}{\alpha^2 r^2}}, \frac{B}{\alpha^2 r^2}, 0
\right),
\ee
where $A$, $B$, and $C$ are constant along each geodesic. We shall choose
$A = e$, $B = l$, and $C = e^2$ for null geodesics, and $A = E$,
$B = L$, and $C = E^2 -1$ for timelike geodesics.  Each geodesic is specified
by the two constants of motion, $(e,l)$ in the case of null geodesics, and
$(E,L)$ in the case of timelike geodesics.
The point of closest approach
to the string is given by $r_c = B/\alpha\sqrt{C}$. The resulting geodesic
equations can be integrated directly to give the following relationship
between $r$ and $\phi$ along the geodesic:
\be
r = \frac{r_c}{cos(\alpha (\phi - \phi_c))},
\ee
where $\phi_c$ is the value of $\phi$ at closest approach.

In order to view these geodesics, we will draw the spacelike slice with
constant $z$ in three different representations:
(1) the ``cone" representation, in which the geodesics are drawn on the
surface of a cone,
(2) the projection of the cone onto two dimensional $r$ and $\phi$ space
referred to here as the ``rubber" representation and
(3) the ``Minkowski spacetime minus a wedge" representation called
the ``paper'' representation.
This nomenclature was chosen as representative of the methods -- to go from
the paper representation to the cone representation, one merely closes the
edges of the missing wedge, thereby lifting the paper from a plane to a
cone.  To go from the paper representation to the
rubber representation, one deforms the paper as though it were rubber,
stretching it so that the edges of the missing wedge are closed while
the rubber sheet remains lying in a plane. Orbits in the rubber
representation are just the orthogonal projections of the geodesics
onto spacelike sections with $t$ constant.
The distinction of the paper representation is that the
missing angle is all accounted for in a wedge pointing in some arbitrary
direction (i.e. by restricting the range of $\phi$), while in the other
two representations, the missing angle is
accounted for uniformly throughout the diagram via the use of the metric
(i.e. by scaling $\phi$).  If carefully applied, all three representations
are equivalent.

To gain further understanding of the shape of the orbits, we
calculate the radius of curvature of the orbits in the rubber
representation. The radius of curvature at a point along the orbit is:
\be
R = r_c \frac{(\alpha^2 sin^2(\alpha (\phi - \phi_c)) + cos^2(\alpha
(\phi - \phi_c)))^{3/2}}{(1 - \alpha^2) cos^4(\alpha (\phi - \phi_c))}.
\ee
The radius of curvature varies with $\phi$ along the orbit, and
reaches its minimum value, $r_c/(1-\alpha^2)$, at closest approach.
For an orbital segment with endpoints $(r_i,\phi_i)$ and $(r_f,\phi_f)$,
we have the relation:
\be
\alpha (\phi_f - \phi_i) = cos^{-1}\left(\frac{r_c}{r_i}\right) +
cos^{-1}\left(\frac{r_c}{r_f}\right),
\label{swept_phi}
\ee
when the segment does in fact have a closest approach. Clearly,
letting the two endpoints go to infinity leads to the result that
$\phi_f - \phi_i = \pi/\alpha$, independent of $r_c$.
Thus the variation in curvature along an orbit occurs in such a way that the
asymptotic deflection from a straight line is $\pi (1 - 1/\alpha)$.
In fact, any two orbits are related by a similarity transformation
corresponding to scaling the closest approach distance, or equivalently,
scaling the radius of curvature.  Any two orbits can be transformed into
each other via rotation and scaling of $r_c$.
This equivalence relation defines an equivalence class of orbits.
For the Minkowski spacetime, this is equivalent to the fact that all
straight lines can be obtained simply by translation and rotation of
any single straight line.

In Figure 1a the cone/rubber/paper representations of
timelike geodesics are presented to demonstrate that
straight lines in the paper representation are bent in the cone and rubber
representations
and that the bending occurs nearer to the closest approach radius for
smaller values of $r_c$.  The diagram shows
that in the rubber representation the curvature near the
closest approach point decreases with increasing $r_c$, yet the angle between
incoming and outgoing asymptotes is independent of $r_c$.  Figure 1b
shows the case for which $\alpha$ is smaller than 0.5; with such
extreme deficit angles, any geodesic of sufficient length will wrap around
upon itself.

It is difficult to find an observer and a source which do not move with
respect to each other over a finite length of time.  The only true test of
the relative motion of a source and an observer is a calculation of the
observed redshift, but the paper representation can easily fool one into
developing an incorrect intuitive notion of comoving observers and sources.
In Figure 1c  we plot the curves corresponding to a geodesic
source and a geodesic observer
that appear in the paper representation to be moving
parallel to each other.  This diagram is closely related to the diagram
used by Kaiser and
Stebbins in discussing the onset of redshift as a string passes by.  In
the rubber representation, it is obvious that
only the asymptotes are parallel.  This representation shows that the
source and observer are never moving
parallel to each other, and implies that the redshift observed between them
does not suddenly turn on as the string passes, but instead always exists
as the string is approached.  Of course the definitive answer is only
achieved by looking at the quantity $u_ak^a$, proportional to the observed
frequency, for a source and observer, each moving along a geodesic and
connected by a segment of a null geodesic, as we shall discuss later.
\begin{figure}[h]
\caption{Figure 1a shows two orbits in the cone/rubber/paper representations
for $\alpha = 0.7$.  Similarly, Figure 1b shows the same orbits for
$\alpha = 0.4$.  Figure 1c shows two orbits for $\alpha = 0.8$ that are,
in the paper representation, apparently parallel.}
\end{figure}

\section{Double Imaging}

Because of the conical nature of the spacetime, there will
be regions where two null orbits will connect the same initial and final
points.  For a given observer any source within a region bounded by the two
null orbits with $r_c$ approaching zero from either side of the string will
produce two images.
This is equivalent to the characteristic double imaging found by
Vilenkin (1984) and later referred to by Kaiser and Stebbins.

Consider some null geodesic segment with initial point $(r_i,\phi_i)$.
The form of the geodesic tangent vector shows that $dr/d\phi$ at $r_i$
can be of either sign, so there are two null geodesics that can leave $r_i$.
In general these two orbits will have distinct closest approach distances,
$r_c$. Let the point where these two orbits cross again be $(r_f,\phi_f)$, and
define the two quantities $\Delta\phi_1$ and $\Delta\phi_2$ to be the angle
swept out along each of the two orbits. Letting $r_{c1}$ and $r_{c2}$ be
the respective distances of closest approach, we can write
equation (\ref{swept_phi}) for each orbit to obtain:
\beqa
\alpha \Delta \phi_1 & = & cos^{-1}(\frac{r_{c1}}{r_i}) +
cos^{-1}(\frac{r_{c1}}{r_f}), \label{invcos1} \\
\alpha \Delta \phi_2 & = & cos^{-1}(\frac{r_{c2}}{r_i}) +
cos^{-1}(\frac{r_{c2}}{r_f}).  \label{invcos2}
\eeqa
{}From each of these equations, the closest approach distance,
$r_{c1}$ or $r_{c2}$, can be found, in the form:
\be
r_{c1}^2 = \frac{r_i^2 r_f^2 sin^2(\alpha\Delta\phi_1)}
{r_i^2 + r_f^2 - 2 r_i r_f cos(\alpha\Delta\phi_1)}.
\ee
To ensure that the two orbits do cross at the points
$(r_i,\phi_i)$ and $(r_f,\phi_f)$, we must require that there is closure in
the angle swept through by the orbits, so that
$\Delta\phi_1 + \Delta\phi_2 = 2\pi$.

For a source moving behind a string, the onset of double imaging occurs
when one of the null geodesics passes near the string with a
vanishingly small distance of closest approach, and the other null geodesic
passes on the other side of the string with a maximum distance of closest
approach.  We choose to take $r_{c1} = 0$
from which we have that $\alpha\Delta\phi_1 = \pi$, so that
$\alpha\Delta\phi_2 = 2\pi\alpha - \pi$. Since
$sin(\alpha\Delta\phi_2) = -sin(2\pi\alpha)$ and
$cos(\alpha\Delta\phi_2) = -cos(2\pi\alpha)$, we find the
maximum distance of closest approach for double imaging:
\be
r_{c \;\; double}^2 = \frac{r_i^2 r_f^2 sin^2(2 \pi \alpha)}
{r_i^2 + r_f^2 + 2 r_i r_f cos(2 \pi \alpha)} .
\ee

When $\alpha = 1$, it follows that $\Delta\phi_1 = \Delta\phi_2 = \pi$ and
hence, $r_{c1} = r_{c2} = 0$; there is only one orbit joining the points
$(r_i,\phi_i)$ and $(r_f,\phi_f)$, as one would expect in a non-conical
Minkowski spacetime.

An observer at $(r_f,\phi_f)$ would measure an angle $\theta$ on the sky
separating the image and the string.  This angle can be found by taking the
dot product of the spatial parts of the null vectors pointing at the
source, $k^i$, and at the string, $\tilde k ^i$,
\beqa
cos \theta & = & \frac{k^i \tilde{k}_i}{\sqrt{k^a k_a}
\sqrt{\tilde{k}^b \tilde{k}_b}} \\
& = & \pm \sqrt{1 - \frac{r_c^2}{r_f^2}} . \label{coslaw}
\eeqa

When the source is in the double image region, there are two different
values of $\Delta\phi$ and two different position angles, $\theta_1$
and $\theta_2$ for the two images.  The
observed angular separation between images is $\theta_s = \theta_1 +
\theta_2$.

Equation (\ref{coslaw}) can be rewritten in the
form $\theta_1 = sin^{-1}(r_{c1}/r_f)$, and
$\theta_2 = sin^{-1}(r_{c2}/r_f)$ for the two image positions.
We will require the derivatives:
\be
\frac{d\theta_1}{dr_{c1}} = \frac{1}{\sqrt{r_f^2 - r_{c1}^2}}
\;\; {\rm and}\;\;
\frac{d\theta_2}{dr_{c1}} = \frac{1}{\sqrt{r_f^2 - r_{c2}^2}}
\frac{dr_{c2}}{dr_{c1}}.
\ee

Using equation (\ref{invcos1}) and (\ref{invcos2}) and the requirement of
closure to obtain $dr_{c2}/dr_{c1}$, we obtain the
derivative of $\theta_s$:
\be
\frac{d\theta_s}{dr_{c1}} = \frac{1}{\sqrt{r_f^2 - r_{c1}^2}}
- \frac{1}{\sqrt{r_f^2 - r_{c2}^2}}
\left(\frac{ 1/\sqrt{r_f^2 - r_{c1}^2} + 1/\sqrt{r_i^2 - r_{c1}^2}}
{ 1/\sqrt{r_f^2 - r_{c2}^2} + 1/\sqrt{r_i^2 - r_{c2}^2}}\right).
\ee

The only time when this expression can vanish is when $r_{c1} = r_{c2}$,
i.e. exact alignment, or when $r_i = r_f$.  Thus in general we can expect
the image seperation to vary as the observer or source moves.

If a point source is moving behind a distant string, a single image appears
in the sky of the observer until that image moves within a critical angle,
$\theta_{double}$, of the string where
\be
\theta_{double} = sin^{-1}\left( \frac{r_i  sin(2\pi\alpha)}
{\sqrt{r_i^2 + r_f^2 + 2 r_i r_f cos(2\pi\alpha)}}\right) .
\ee
At this point the two images appear separated by the angle $\theta_s =
\theta_{double}$, one image lying very close to the string.  As the source
continues to move across the sky, both images move in the same direction in
such a way that their angular separation, $\theta_s$, will vary.  When the
leading image is at the critical angle, the separation angle between the
images has returned to $\theta_{double}$.  A moment later, only one image
exists, moving away from the string.

\section{Redshift and CMB anisotropy}

In order to consider the effects of a cosmic string on the microwave
background, we will model the microwave background as a continuous collection
of point sources.  Initially, we will require these sources to be at rest
with respect to the string, and to each other.  For any one of these
sources, the four-velocity is $u^a = (-1,0,0,0)$.  We introduce a general
timelike observer with four velocity
\be
u^a = \left(-E, \pm\sqrt{E^2 -1
- \frac{L^2}{\alpha^2r^2}},\frac{L}{\alpha^2r^2},0\right).
\ee
If the null geodesic which joins the source and observer has a tangent vector
given by
\be
k^a = \left(-e, \pm\sqrt{e^2 - \frac{l^2}{\alpha^2r^2}},
\frac{l}{\alpha^2r^2},0\right),
\ee
then the redshift is given by:
\beqa
1 + z & = & \frac{\nu_{source}}{\nu_{obs}} = \frac{(u^ak_a)_{source}}
{(u^bk_b)_{obs}}\\
 & = & \frac{-1}{-E \pm \sqrt{E^2 - 1 - L^2/(\alpha^2 r_f^2)}
\; \sqrt{1 - {r_c^2}/{r_f^2}} \pm r_c L/(\alpha r_f^2) }.
\eeqa

In terms of the image position angle on the observer's sky, $\theta$, this
is:
\be
1 + z = \frac{1}{E + B\,cos\theta + C\,sin\theta},
\ee
where $B= \sqrt{E^2 - 1 - L^2/(\alpha^2 r_f^2)}$ and
$C = L/(\alpha r_f)$ are constants depending only upon the motion and
position of the observer.

The direction of the peak of redshift $\theta_0$, found by setting
$dz/d\theta = 0$, is given by $tan\theta_0 = C/B$.  Introducing a new angle
$\theta^\prime = \theta - \theta_0$ centered about $\theta_0$ gives:
\be
1 + z = \frac{1}{E + B_0^\prime \,cos\theta^\prime},
\ee
where $B_0^\prime = B_0 (cos\theta_0+tan\theta_0\,sin\theta_0)$.

This is the form of a simple dipole due to the observer's motion through
the cosmic radiation.  This result contrasts with Kaiser and Stebbin's
moving wedge model in which a jump discontinuity is observed at the string.
The dipole predicted by this method is essentially due to the motion of the
observer and does not reveal any characteristics of the string.  The same
result will hold for a collection of ``comoving" sources:  if we modify our
model of the microwave background by introducing a moving collection of
sources, we will  still find a dipole result.  If the collection of sources
is moving such that there is no redshift between each pair of sources, and
yet there is a redshift between one of the sources and the string, then an
observer can be boosted to have no redshift with respect to that one
source.  Obviously, the observer must have no redshift with respect to the
other sources as well.  Thus, any redshift seen by the observer can be
boosted away: any redshift is simply due to the motion of the observer.

Naturally, it would be possible to choose a collection of moving sources
that would be observed as a non-dipole pattern of redshifts, but the
fundamental result is still the same:  the only redshift
which exists on this spacetime is a Doppler shift.  There is no redshift
due to a time varying potential as used in ``the moving gravitational
lens method'' adopted by Kaiser and Stebbins.  The spacetime is
Minkowskian everywhere except at the vertex of the cone.

\section{Conclusion}

The implications of this work for the study of cosmic strings flow from two
points:  it is unknown how  to move a topological structure and understand
the back-reaction of the rest of the universe; and photons
moving on a conical spacetime experience only Doppler shift, not
gravitational redshift.
Until we are capable of understanding topological dynamics, we must treat
cosmic strings in a fully relativistic manner, taking seriously the conical
nature of the spacetime.
It is hoped that eventually the dynamics of a cosmic string will be better
understood so that problems such as the evolution of
cosmic strings during an early universe phase transition, and the motion of
cosmic strings with respect to each other can be treated with more rigour.
Since strings are created before the last scattering surface, it is
possible that the changes in topology occurring during the formation of strings
could cause density perturbations that would present themselves as
temperature fluctuations in the microwave background.
Until it is possible to understand such mechanisms, we must work with the
observational consequences of microwave background radiation propagating in the
vicinity of a cosmic string.  The background will be largely homogeneous and
isotropic, and will contain some small temperature variations.  The
question that must be answered is whether or not further inhomogeneities are
introduced by the presence of a cosmic string.
Kaiser and Stebbins
developed the presently accepted means of answering this question with the
``Minkowski minus a wedge representation'' of a moving string and found
that the string introduced further temperature fluctuations.
We have taken the conical
spacetime and treated it as any other spacetime in general relativity: it
is the arena in which sources and observers move.
Using this method, we have shown that the string does not introduce further
fluctuations in the temperature of the microwave background.
The temperature of the cosmic microwave background would be entirely isotropic
except for a dipole due to the motion of the observer.

\acknowledgments{
This work has been supported by the Natural Sciences and Engineering
Research Council of Canada through a Postgraduate Scholarship (F.~R.~M.),
a Postdoctoral Fellowship (J.~H.~P.) and an operating grant (C.~C.~D.).
}

\end{document}